\documentclass{aa}
\usepackage{graphicx}

\begin{document}
\title{Long Period Variables in NGC~5128: II.\@ Near-IR
properties\thanks{Based on
observations collected at the European
Southern Observatory, Paranal, Chile, within the Observing Programmes
63.N-0229, 65.N-0164, 67.B-0503, 68.B-0129 and 69.B-0292, and at La Silla
Observatory, Chile, within the Observing Programme 64.N-0176(B).}}

\author{M. Rejkuba\inst{1}
    \and D. Minniti\inst{2}
    \and D.R. Silva\inst{1}
    \and T.R. Bedding\inst{3}}

\offprints{M. Rejkuba}

\institute{European Southern Observatory, Karl-Schwarzschild-Strasse
           2, D-85748 Garching, Germany\\
           E-mail: mrejkuba@eso.org, dsilva@eso.org
  \and Department of Astronomy, P. Universidad Cat\'olica, Casilla
        306, Santiago 22, Chile\\
        E-mail: dante@astro.puc.cl
  \and School of Physics, University of Sydney 2006, Australia\\
        E-mail: bedding@physics.usyd.edu.au}

\date{07 July 2003 / 29 August 2003}
\authorrunning{Rejkuba et al.}
\titlerunning{LPVs in Cen A}

\abstract{Long Period Variable stars are ubiquitous among the bright red
giant branch stars in NGC~5128. Mostly they are found
to be brighter than the tip of the first ascent giant branch with magnitudes
ranging from about $K=19$ to $K=21.5$. They have periods
between 155 and 1000 days and K-band amplitudes between 0.1 and 2 mag,
characteristic of semi-regular and Mira variables. We
compare the colors, periods and amplitudes of these variables with those
found in old stellar populations like Galactic globular clusters and
Galactic bulge as well as with intermediate-age Magellanic Cloud long period
variables.
The population of stars above the tip of the red giant branch (RGB)
amounts to 2176 stars
in the outer halo field (Field~1) and  6072 stars in the inner halo
field (Field~2). The comparison of the luminosity functions
of the Galactic bulge, M31 bulge and NGC~5128 halo fields shows an 
excess of bright AGB stars extending to $M_K\simeq -8.65$. 
The large majority of these sources belong to the
asymptotic giant branch (AGB)
population in NGC~5128. Subtracting the foreground
Galactic stars and probable blends, at least 26\% and 70\%
of AGB stars are variable in Fields~1 and 2, respectively.
The average period of NGC~5128 LPVs
is 395 days and the average amplitude 0.77 mag. Many more short period
Miras are present in Field~2 than in Field~1 indicating a difference in the
stellar populations between the two fields. 
Period and amplitude distributions and
near-IR colors of the majority of LPVs in NGC~5128
are similar to the Galactic bulge variables. However, some $\sim 10$\%
of LPVs have periods longer than 500 days and thus probably more massive,
hence younger, progenitor stars. A few carbon star candidates are 
identified based on their red $J-H$ and $H-K$ colors.
\keywords{Galaxies: elliptical and lenticular, cD --
          Galaxies: stellar content --
          Stars: fundamental parameters --
          Galaxies: individual: NGC~5128}
}

\maketitle

%
%
\section{Introduction}

In intermediate-age populations
($\sim 1 - 5$~Gyr old) numerous bright asymptotic giant
branch (AGB) stars are located above the tip of the RGB.
However, bright stars have also been
detected above the tip of the RGB among old populations like
Galactic globular clusters with ${\rm [Fe/H]}\la -1.0$~dex
and in the Galactic bulge (Frogel \&
Elias~\cite{frogel&elias88}, Guarnieri et al.~\cite{guarnieri+97},
Momany et al.~\cite{momany+03}),
implying the presence of bright AGB stars
in metal-rich and old populations. All of the bright
giants above the RGB tip in globular clusters seem to be long period
variables (LPVs; Frogel \& Elias~\cite{frogel&elias88}, Frogel
\& Whitelock~\cite{frogel&whitelock98}).
The frequency of LPVs in old metal-rich globular clusters of the MW
and in the Bulge has been studied by
Frogel \& Whitelock~(\cite{frogel&whitelock98}).
Old populations of lower metallicity are known not to
have AGB stars brighter than the RGB tip.

The brightest LPVs in old stellar populations like the
Galactic globular clusters and the old disk in the
Milky Way reach  $M_K=-8$ (Mennessier
et al.~\cite{mennessier+01}, Feast et al.~\cite{feast+02},
Momany et al.~\cite{momany+03}).
The higher mass (hence younger) LPVs in the LMC and in the disk of our
galaxy can be more luminous, reaching $M_K=-9.4$ (Hughes
\& Wood \cite{hughes&wood90}, Mennessier et al.~\cite{mennessier+01}).

There exists a long standing debate about the presence of intermediate-age
population in dwarf elliptical galaxies and spiral bulges. On the basis of
the Wide Field Camera 1 (WFC1) HST data, Holtzman
et al.~(\cite{holtzman+93}) and Vallenari et al.~(\cite{vallenari+96})
concluded that the majority of stars in the
Galactic bulge are of intermediate age. Other studies indicated mainly old
metal-rich population (e.g. Frogel et al.~\cite{frogel+90}, Tiede et
al.~\cite{tiede+95}). Newer deep optical and near-IR
CMDs suggest that the Bulge is old without even a trace of an
intermediate-age population (e.g. Ortolani
et al.~\cite{ortolani+95}, Feltzing \& Gilmore~\cite{feltzing&gilmore00},
Zoccali et al.~\cite{zoccali+03}). In the M32 dwarf elliptical galaxy
the situation might be different, given the presence of an extended giant
branch reaching $M_K=-8.7$ ($M_{bol}=-5.5$), indicative of an
intermediate-age ($\sim 4$~Gyr) population, which has been revealed by
near-IR photometry by Elston \& Silva (\cite{elston&silva92})
and Freedman (\cite{freedman92}). However, it is of interest to note
that in the deep optical $VI$ photometry of M32 with HST, no optically
bright AGB stars have been found (Grillmair et al.~\cite{grillmair+96}),
which might be expected due to large bolometric corrections.

In the gE NGC~5128, Soria et al.~(\cite{soria+96}), suggested the
presence of up to 10\% bright AGB stars belonging to an
intermediate-age population,
based on a $VI$ HST CMD\@.   Marleau
et al.~(\cite{marleau+00}) made a similar suggestion based on $JH$ NICMOS data.
Harris et al.~(\cite{harris+99,harris&harris00}), on the contrary,
do not find any bright AGB stars in their $VI$ CMDs of two halo fields
in NGC~5128. As mentioned above, $V$ and $I$ bands are not
very sensitive indicators of these cool giants and thus some of them might
have been confused with first ascent red giants,
foreground stars and few stellar blends or stars
with larger photometric errors. In the two fields observed
in $V$ and $K$-bands with VLT (Rejkuba et al.~\cite{rejkuba+01}), a
large number of bright AGB stars have been detected extending up
to bolometric magnitude of $-5$. Most of these stars have been found to
be variables in our long term monitoring programme with ISAAC at VLT.
The LPV catalogue has been published by
Rejkuba et al.~(\cite{rejkuba+03}).
Here we analyse near-IR properties
of these variables and compare them with the LPV population found in the
Magellanic Clouds and in the Galactic bulge, with the aim of
constraining the contribution of intermediate-age stars to the
NGC~5128 halo population.

%
%

\section {The Catalogue}
\label{data}

The LPV variables catalogue in NGC 5128 has been compiled and published
by Rejkuba et al.~(\cite{rejkuba+03}). The reader is referred to that paper
for the details on the data reduction and the determination of periods.
In summary, the observations consist of single-epoch 1-hour exposures in
$J_s$ and $H$-bands and multi-epoch photometry in $K_s$-band 
obtained with ISAAC
near-IR array at ESO Paranal Observatory with UT1 VLT\@.
The two fields observed are located $\sim 17'$ north-east (Field~1)
and $\sim 9'$ south (Field~2) from the center of NGC 5128.

$K_s$-band observations span a time interval of 1197 days,
from April 1999 till July 2002, with 20
epochs in Field~1 and 23 in Field~2. One
additional 45 min observation of Field~2 in $K_s$-band with SOFI at the NTT at
La Silla Observatory brings the total number of epochs to 24 for that field.
It should be noted that SOFI near-IR array is a scaled version of ISAAC and
that the $K_s$-band filters at the two instruments are identical.

The photometry of this homogeneous data set has been performed with
DAOPHOT and ALLFRAME software (Stetson~\cite{stetson94}). The 50\%
completeness limits in $K_s$ 
and $H$-bands are at $22.5$ for Field~1 and $21.5$ for
Field~2. In $J_s$-band 50\% completeness
is achieved at $Js=23.25$ for Field~1 and $Js=22.5$ for Field~2.
To all $K_s$-band magnitudes
from the catalogue (Rejkuba et al.~\cite{rejkuba+03}) we have subtracted
0.1 mag.
This was necessary in order to match the observed colors of foreground
dwarf and giant
stars with that of Bessell \& Brett (\cite{bessell+brett88}) fiducials
(see also Sect.~\ref{color-color-sect} and Rejkuba~\cite{rejkuba03}).
We have re-checked the photometric calibration and most probably this
0.1 offset in $K$-band is due to an error in the aperture correction.
It should be noted that the aperture corrections in
such crowded fields are rather uncertain and very difficult to measure
accurately.

There are total of 15574 sources in Field~1 and 18098 in Field~2 that were
detected in at least 3 $K_s$-band epochs. Restricting the detection to
$J_s$, $H$ and 3 $K_s$-band 
epochs the total number of objects is 13111 and 16434
in Fields 1 and 2, respectively. Among these there
are more than 1500 variable stars. For 1046 red variables, with at least 10
$K_s$-band measurements, periods and amplitudes were determined with
Fourier analysis and then refined with non-linear fitting of sinusoidal
function:
\begin{equation}
K(t) = A \cos \left( 2 \pi \frac{(t - t_0)}{P} \right) +
B \sin \left( 2 \pi \frac{(t - t_0)}{P} \right) + K_0
\end{equation}

In the final catalogue (Rejkuba et al.~\cite{rejkuba+03})
periods of variable stars are average values determined from
the Fourier analysis and the non-linear sine-curve fitting algorithms, except
in the cases where a visual inspection of the light curve clearly preferred
one of the two cases. In most cases the two periods were nearly equal (see
Rejkuba et al.~\cite{rejkuba+03} for simulations and detailed discussion of
period accuracy). Amplitudes, $a = 2\times \sqrt{A^2+B^2}$, and
average $K$-band magnitudes
are derived from sine-curve fitting.

A close inspection of periods of LPVs lying away from the Mira PL relation
indicated a problem of aliasing periods of $\sim 1/2$ and $\sim 1$ year.
For a total of 39 variables the most significant period in
Fourier periodogram was shorter, approximately 1/2 year,
but they could be equally well fit by twice as long periods.
These are listed by Rejkuba (\cite{rejkuba03}; Tab.~1). Other 7
variables have possible shorter periods and for 6 more LPVs improved
amplitudes and mean magnitudes have been determined, but their
periods remained unchanged to within 3\%.
In the following analysis periods, amplitudes and mean magnitudes
from the LPV catalogue (Rejkuba et al.~\cite{rejkuba+03}) will be used
except for those
listed in Tab.~1 of Rejkuba (\cite{rejkuba03}). However, the results of
our analysis are not influenced by this choice.

%
%

\section{LPVs in near-IR color-magnitude diagrams}
\label{NIR CMDs}

\begin{figure*}
\centering
\includegraphics[width=8.9cm,angle=0]{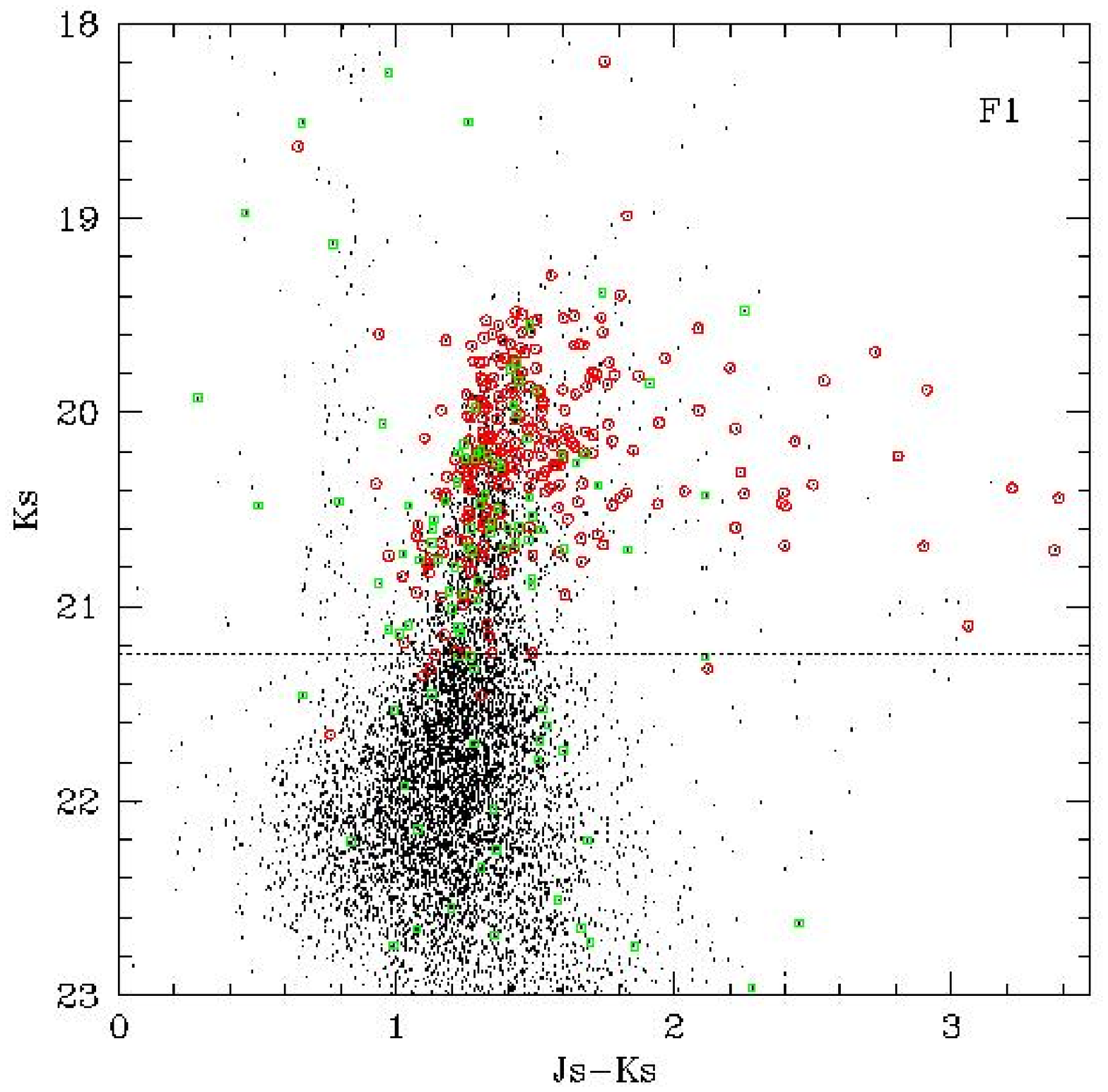}
\includegraphics[width=8.9cm,angle=0]{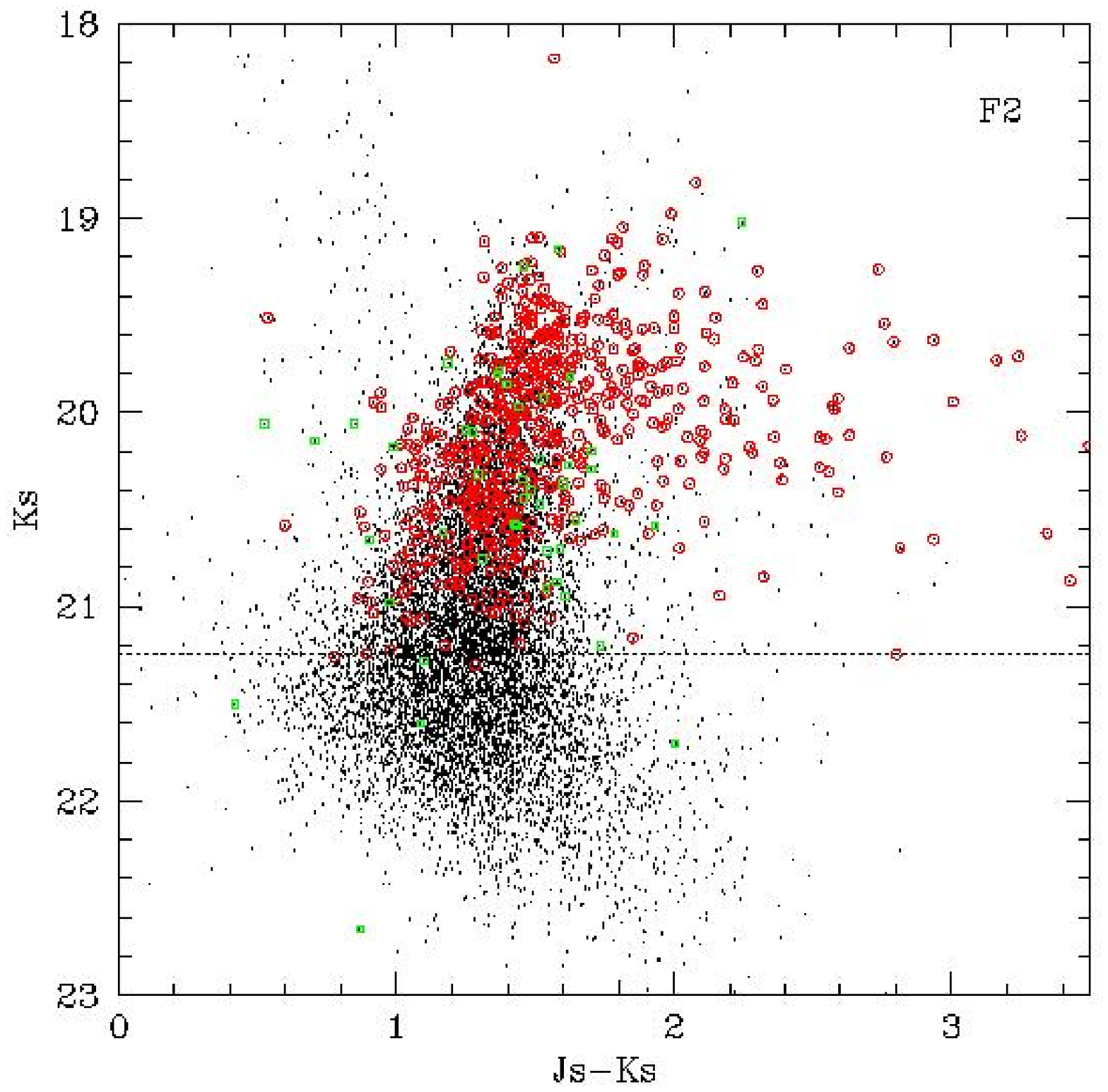}\\
\includegraphics[width=8.9cm,angle=0]{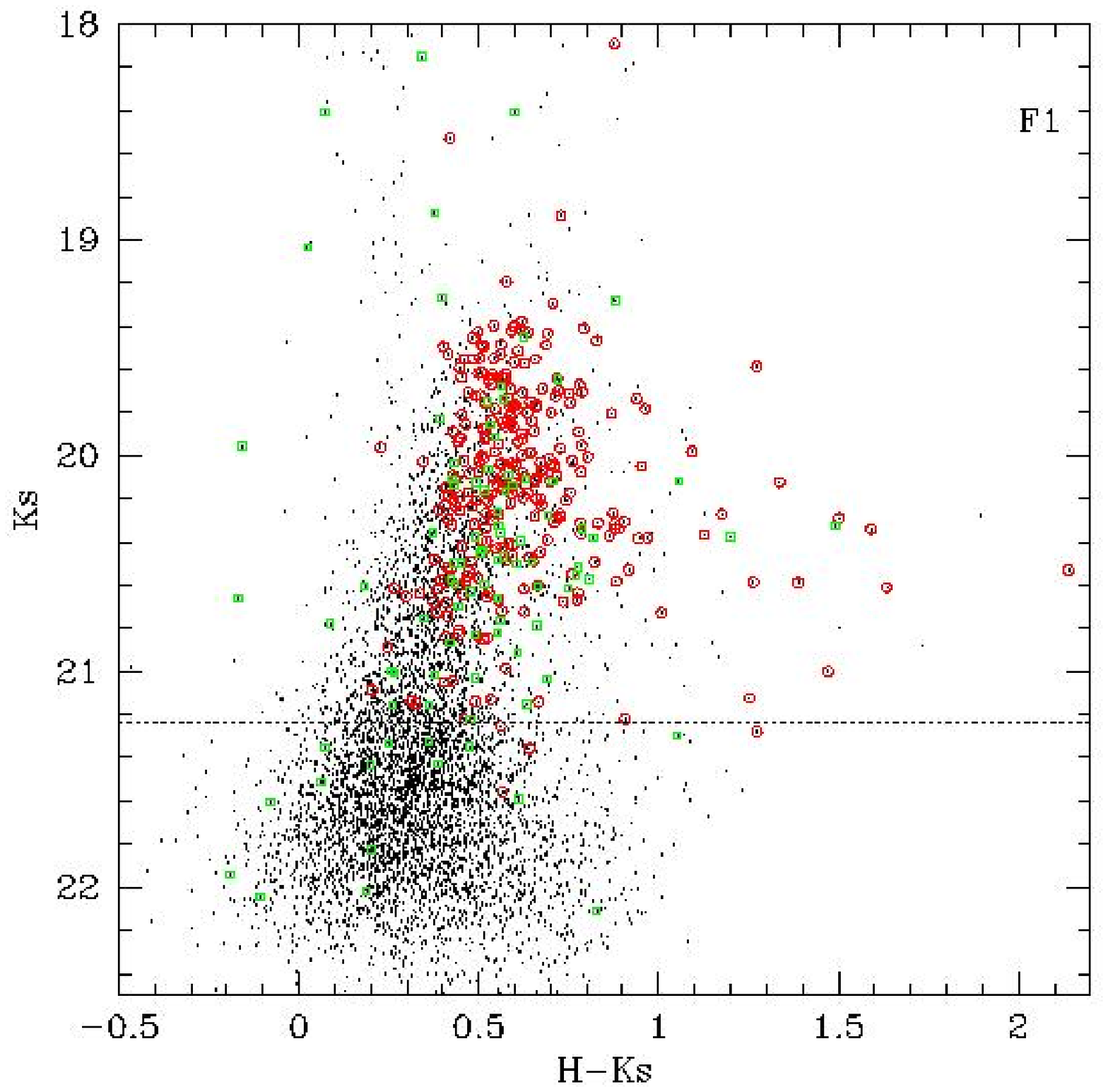}
\includegraphics[width=8.9cm,angle=0]{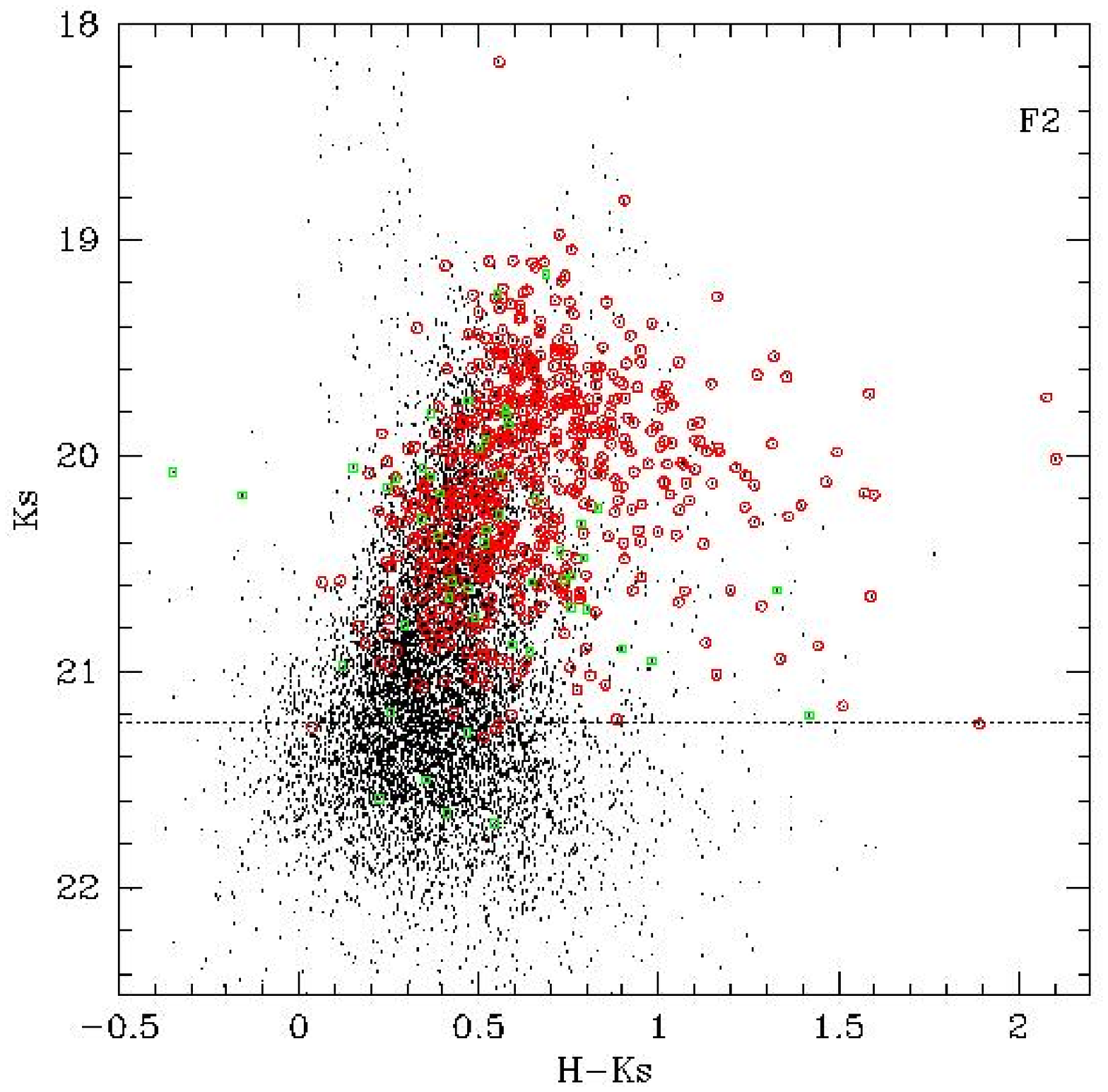}
\caption[]{Top panels:
	$K_s$ vs. $J_s-K_s$ color-magnitude diagrams for all the stars in
	Field~1 (left) and Field~2 (right) is shown with black dots.
	Bottom panels:
	$K_s$ vs. $H-K_s$ color-magnitude diagrams for Field~1 (left) and
	Field~2 (right) stars.
	Variable stars with reliably determined periods
	(significance parameter from Fourier analysis $<0.7$) are
	indicated with red circles and green boxes indicate those with
	less reliable periods (significance~$\geq 0.7$). Dotted
	line at $K_s=21.24$ shows RGB tip magnitude.}
\label{cmdlpv}
\end{figure*}

In Fig.~\ref{cmdlpv} we show $K_s$. vs. $J_s-K_s$ color-magnitude diagrams
(CMDs) for
all the stars detected in at least 3 $K_s$-band epochs as well as in $J_s$
and $H$-band images that had ALLFRAME photometric errors smaller than 0.25 mag.
Field~1 CMD is displayed in the left and Field~2 in the right panel.
Long period variable stars with reliably determined periods
(significance parameter from Fourier analysis $<0.7$) are
indicated with red circles and green boxes indicate those with
less reliable periods (significance~$\geq 0.7$). In the following
analysis only those variables with better determined periods
(significance~$< 0.7$) will be used. These are 280 LPVs in Field~1 and 617
LPVs in Field~2.

Dotted line at $K_s=21.24$ shows RGB tip magnitude measured from
the $K$-band luminosity
function (Rejkuba \cite{rejkuba03}).
There is a plume of foreground Galactic stars that are bluer
and well separated from the red giants in NGC~5128. According to field
simulations from Marigo et al.\ (\cite{marigo+03}) they are mainly disk stars.
In particular, old disk turn-off stars are
expected to be distributed mainly at $(J_s-K_s) \simeq 0.36$, Galactic
RGB and red clump stars have colors around $(J_s-K_s) \simeq 0.65$
and low mass dwarfs with $M \leq 0.6$~M$_\odot$ are found around
$(J_s-K_s) \simeq 0.85$. The fact
that they are distributed in vertical sequences is entirely due to a range of
distances that they span. No Galactic foreground contamination is expected
at $(J_s-K_s) \geq 1.0$, where red giants in NGC~5128 are located.
The equivalent $K_s$ vs. $H-K_s$ CMDs are shown in the two lower panels of
Fig.~\ref{cmdlpv}. Here there
is some overlap between the regions where foreground sources and red giants
in NGC~5128 are located. It should be noted from these CMDs that, due to
the much higher density of stars in Field~2 and slightly worse
average seeing during the observations and higher total luminosity sampled,
the limiting magnitude is brighter and the
photometric scatter is much larger.

The $K_s$-band magnitude used to compose $J_s-K_s$ and $H-K_s$
colors in Fig.~\ref{cmdlpv} is a single epoch measurement obtained closest in
time to the respective $J_s$ or $H$-band epoch.
The time separation between the two is less than 10 days.  In contrast,
the $K_s$ magnitude plotted on the y-axis is a mean
value from all the epochs and it well represents mean magnitudes of AGB stars.
If a mean $K_s$ magnitude is used for $J_s-K_s$ and $H-K_s$ color, a
much larger scatter in color is observed in the part of the CMD where
variable stars are found.

%
%

\section{Amplitude distribution}

\begin{figure}[t]
\centering
\includegraphics[width=7cm,angle=270]{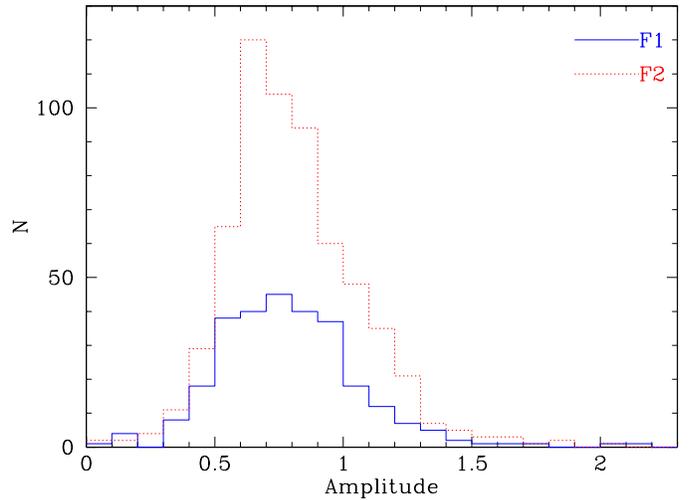}
\caption[]{Histogram showing the amplitude
		distribution of LPVs in Field~1 (solid line) and 2 (dotted
		line).}
\label{ampl_hist}
\end{figure}

Amplitudes of the LPVs in NGC~5128 were determined as peak-to-peak magnitude
differences of the best fitting sinusoid light-curve.
Fig.~\ref{ampl_hist} shows the amplitude distribution of all the LPVs with well
determined periods. The majority of variables have
$K$-band amplitudes of around 0.7 mag, with median value of 0.77
in both Fields. The $K$-band amplitudes for LPVs
in LMC (e.g.\ Wood et al.~\cite{wood+83}) and in the Solar neighbourhood
(e.g.\ Whitelock et al.~\cite{whitelock+00}) range between 0.1 and 1.2 mag.
Only a few LPVs in the Sgr I field in the Bulge have $K$-band amplitudes larger
than 1 mag with the largest amplitude of 2.5 mag.

\begin{figure}[t]
\centering
\includegraphics[width=7cm,angle=270]{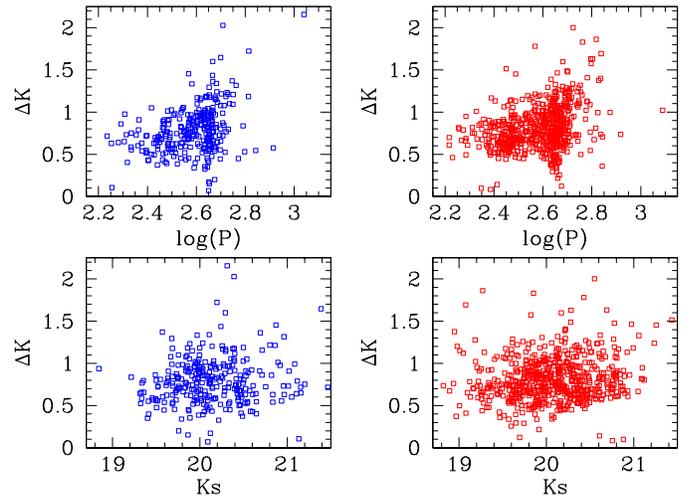}
\caption[]
	{Amplitudes as a function of periods (top) and
	mean magnitudes (bottom) for Field~1 (left)
	and 2 (right).}
\label{ampl_magP}
\end{figure}

The dependence of the amplitudes on the mean magnitudes and on the periods
is shown in Fig.~\ref{ampl_magP}.
As was also found for Galactic Mira-like LPVs
(Whitelock et al.~\cite{whitelock+00}), there
is a very weak correlation in the sense that longer period variables tend
to have larger amplitudes. Similarly Glass et al.~(\cite{glass+95}) found
that the large-amplitude stars have periods longer than 350 days
in the Sgr I field in the Galactic bulge.

There are only 32 (out of 897) variables with amplitudes smaller
than $\Delta K<0.4$. Such small amplitude variables usually tend to 
have fainter
magnitudes. Our catalogue is much less complete for these variables. While
some of them may have been detected as variable stars, no reliable periods
could be measured for them. Semi-regular variables have smaller amplitudes
making the detection of periodic variability even more difficult with a
small number of observations. Additionally, some of them might not even be
detected as variable due to combination of small amplitudes and
photometric uncertainty. They might comprise the large majority of
the ``non-variable'' AGB stars detected in Field~1 (see also the next section).

\section{Luminosity function}
\label{LF_K}

\begin{figure}[t]
\centering
\includegraphics[width=7cm,angle=270]{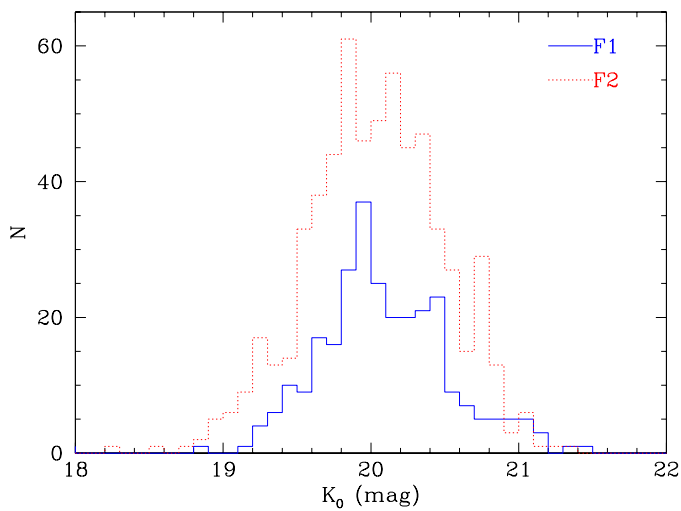}
\caption[]
	{$K$-band luminosity function for all variables with well
	established periods (Fourier analysis significance $<0.7$)
	in Field~1 (solid line) and Field~2 (dotted line).
	}
\label{lfkvar}

\end{figure}
\begin{figure}[t]
\centering
\includegraphics[width=7cm,angle=270]{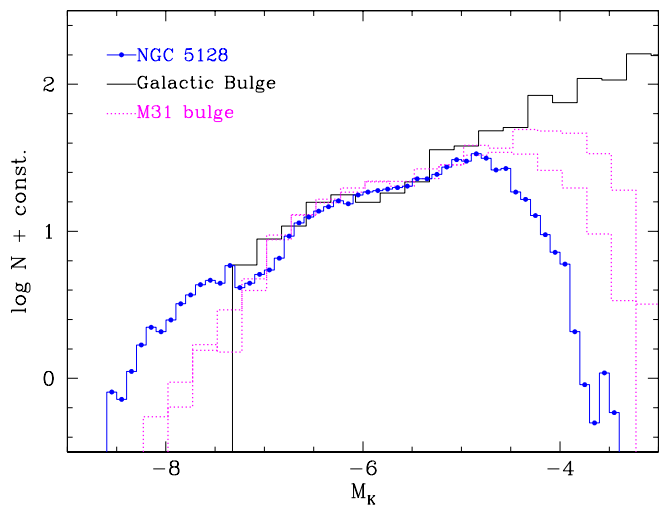}
\caption[]
	{Comparison of the $K$-band luminosity functions of Field~1 in NGC~5128
	halo (beaded blue histogram) with those of the Galactic bulge 
	(solid black histogram) and of two fields in the bulge of M31 
	(dotted magenta histograms). All luminosity functions have been 
	normalized in the range $-6.5<M_K<-5.5$. The Galactic bulge data
	are from Zoccali et al.~(\cite{zoccali+03}) and the
	two M31 bulge luminosity functions are from Stephens et al.\
	(\cite{stephens+03}; fields F3 and F4).
	}
\label{compareLFs}
\end{figure}

The luminosity functions for all the variables with well determined periods
is shown in Fig.~\ref{lfkvar} with solid line for Field~1 and dotted line for
Field~2 LPVs. The luminosity functions in the two fields are very similar.

A detailed calculation of the
expected number of LPVs in a $\sim 10$~Gyr old stellar population
according to Renzini's (\cite{renzini98}) work yields
250-500 LPVs in Field~1 and between 350 and 1500 LPVs in
Field~2 (Rejkuba~\cite{rejkuba03}). The large
uncertainty of the expected number of LPVs in Field~2 comes from
highly uncertain total luminosity of this field due to higher number
of blends and difficulty to correctly estimate the sky
background. The numbers of expected LPVs are in very good agreement with
the 437 and 709 LPVs with measured periods in Fields 1 and 2,
respectively. This would suggest that the majority of the LPVs belong
to an old and metal-rich population similar to that of
our Galactic bulge.

Taking into account the foreground Galactic contamination and a
probable number of blends, LPVs account for at least 26\% and 70\%
of the extended giant branch population in
Fields 1 and 2, respectively (Rejkuba \cite{rejkuba03}).
The completeness of the LPV catalogue in the two fields is rather
similar as a  function of period and amplitude and thus there seems
to be an excess of non-variable bright AGB stars in Field~1. At least
some 1150 stars brighter than the RGB tip cannot be accounted for by
LPVs, foreground stars or blends in Field~1. In Field~2 this number is
an order of magnitude lower and can well be accounted by the incompleteness
of the variable star catalogue. Basically all the constant luminosity 
AGB stars in Field~1 are fainter than $K_s \sim 19.8$.
As mentioned above they could be semi-regular variables with
amplitudes up to about $\Delta K \la 0.3$~mag, the limit at which the LPV
catalogue completeness starts to drop rapidly from $\sim 70\%$. Virtually no
variables with $\Delta K \sim 0.1$ or smaller are expected to be detected
according to the completeness simulations (Rejkuba et al.~\cite{rejkuba+03}).
In Field~2 small amplitude variables could be hidden among blends.
It should be noted that in the Galactic bulge and in old and metal-rich
globular clusters, no constant luminosity 
stars are expected to be found above the
RGB tip (Frogel \& Whitelock \cite{frogel&whitelock98}, Glass \& Schultheis
\cite{glass+schultheis02}). New, detailed searches for variable stars in the
LMC indicate that the large majority of the stars above the
RGB tip, as well as many fainter than the RGB tip, are variable
(Wood \cite{wood00}, Ita et al.~\cite{ita+02}, Kiss \& Bedding
\cite{kiss&bedding03}).

The positions of the brightest LPVs have been carefully checked on the best
seeing images. They show that the three brightest Field~1 variables are
much brighter than expected, due to blending. One of them, with
$<K>\simeq18.1$ and $J_s-K_s=1.75$,
is actually blended with a background galaxy and the very blue bright LPV at
$J_s-K_s=0.55$ is contaminated by a bright neighbouring foreground
star. The brightest 6 LPVs in Field~2 do not seem to suffer any blending.
 
The comparison of the Field~1 $K$-band luminosity function with 
the luminosity functions of the Galactic bulge (Zoccali et 
al.~\cite{zoccali+03}) and of two bulge fields of the M31 
(fields F3 and F4 from Stephens et al.~\cite{stephens+03}) 
is shown in Fig.~\ref{compareLFs}. 
All luminosity functions have been normalized in the range $-6.5<M_K<-5.5$.
Taking into account the blends, NGC~5128 Field~2 luminosity function 
is similar to that of the Field~1 (see also Rejkuba~\cite{rejkuba03}). 
The tip of the AGB is around $M_K=-7.5$ in the Galactic bulge. However,
Mira variables are known in Baade's Window and in Sag I fields that can
reach $M_K=-8$ (Glass et al.~\cite{glass+95}, 
Shultheis \& Glass \cite{schultheis+glass01}), similar to 
the AGB tip observed in the M31.
In NGC 5128 there are many more AGB stars
and the tip of the AGB is more than half a magnitude brighter, 
clearly implying the presence of the intermediate age 
population in the halo. 
The tip of the AGB, measured
from the average $K_s$-band magnitude of the LPVs, is
$K_0=19.3$.
Rejkuba (\cite{rejkuba03}) has measured distance modulus of NGC~5128 of
$27.92 \pm 0.19$~mag, implying that the brightest AGB variables in NGC~5128,
which are not brightened by blending, have $M_K=-8.65$.
Bolometric corrections were calculated
from Montegriffo et al. (\cite{montegriffo+98}) and the brightest AGB
variables in Field~1 have $M_{bol}=-5.3$, while there are few Miras 
in Field~2 that reach 
reach $M_{bol}=-5.7$~mag. The brightest LPV in Field~2 has actually
$M_K=-9.78$ and $M_{bol}=-6.43$. Its red colors suggest that it is a
long period variable. Its period is 696 days and it lies $\sim 0.8$ mag
above the Mira PL relation (Rejkuba \cite{rejkuba03}). Its amplitude is
0.36 mag, rather low for its long period. Unless it is a blend
of two perfectly aligned stars (hence not visible on $0\farcs36$ image), it
is a good candidate for a hot-bottom burner
(Bl\"ocker \& Sch\"onberner~\cite{bloecker+schoenberner91}).

\section{Period distribution of LPVs}
\label{period_amplitude}

\begin{figure}
\centering
\includegraphics[width=7cm,angle=270]{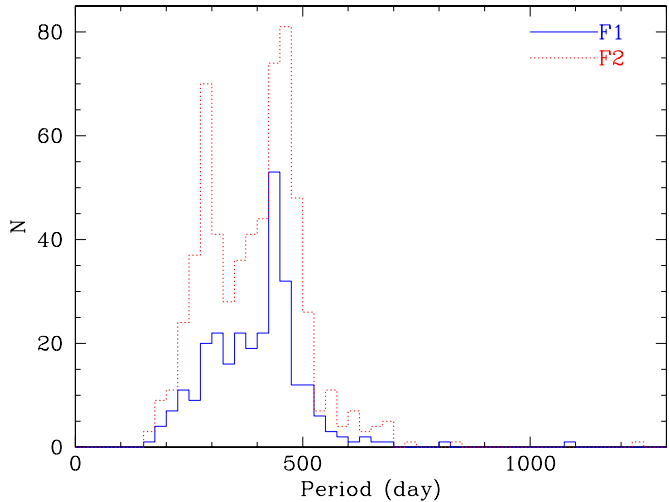}
\caption[]{Histogram showing the period
		distribution of LPVs in Field~1 (solid line) and 2 (dotted
		line).}
\label{Nperiod}
\end{figure}

\begin{figure}
\centering
\includegraphics[width=7cm,angle=270]{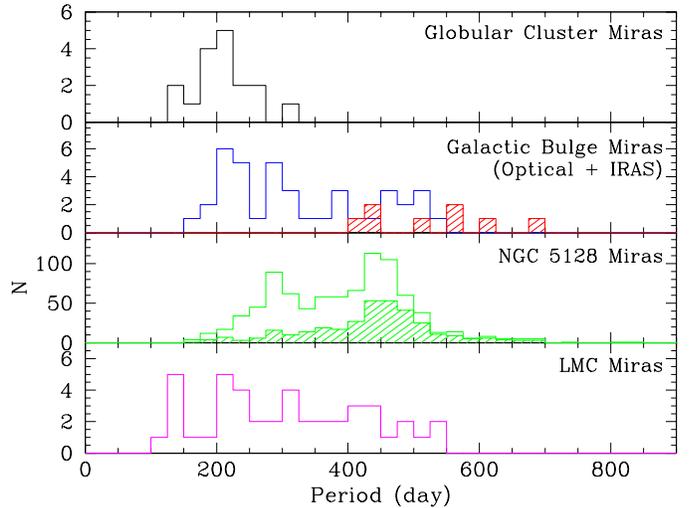}
\caption[]{Comparison of period distributions for Miras in globular
	clusters (Feast et al.~\cite{feast+02}), in the Galactic 
	bulge SgrI field (Glass et al.~\cite{glass+95}) and in the LMC
	(Cioni et al.~\cite{cioni+01}) with NGC~5128 Miras in both fields
	studied in this work. Optically discovered Miras in the SgrI field
	in the Galactic bulge are 
	shown as open histogram, while the 10 IRAS sources 
	identified in the same field are plotted with 
	shaded histogram. Miras ($\Delta K>0.4$) in NGC~5128 with $J_s-K_s$
	colors redder than $J_s-K_s>1.6$ are shown with shaded histogram.} 
\label{compareP}
\end{figure}

The distribution of periods for all the variables with good periods is
shown in Fig.~\ref{Nperiod}. The distributions for Field~1 and 2 are similar
at the long period end, but there is a clearly larger percentage of stars with
periods in the range of 250-300 days in Field~2. This is the same period
range as the old and metal-rich LPVs in Galactic globular clusters. Since the
completeness of the variable star catalogue in this period range is
of the order of 80-90\%, 
the larger number of short period variables in
Field~2 indicates differences in the stellar populations.

The shortest periods are 160 and 155 days, respectively.
The longest reliable periods are 1097 in
Field~1 and 1229 in Field~2. Due to the limited time interval spanned
by the observations, amounting to 1197 days, most of the variables with
well determined periods, hence observed through at least one to two
periods, have periods shorter than 600-700 days. If there is a
larger population of variables with periods in excess of $\sim 700$ days,
only those with larger amplitudes would be detected. The steep drop
in the period distribution present already around P=500 days
indicates that there is no significant population of such very long period
variables. In the present distribution, LPVs with periods longer than 500 days
make 10\% and 11\% 
of the population in Fields~1 and 2, respectively.
For periods longer than 550 days, the contribution to the distribution
drops to 4\% and 6\%.
The average period of 897 LPVs with well determined periods is 395 days.
For 52 variables, periods were revised by Rejkuba~(\cite{rejkuba03}).
If instead, for all the stars periods from the LPV catalogue
(Rejkuba et al.\ \cite{rejkuba+03})
are used, the average period is 388 days.

There are 58 variables in Field~1 and 25 in Field~2 for which
best fitting periods were shorter than 150 days. However, all of these
period determinations are less secure. Some short period
variables, belonging to Cepheid class are expected to be found in
Field~1 where a conspicuous recent star formation is observed
(Mould et al.~\cite{mould+00}, Rejkuba et al.~\cite{rejkuba+01}). Their
colors would be bluer than those of Miras and semi-regulars. Most of the
blue variables detected in our fields are unfortunately blended with
foreground stars.

The period distribution of LPVs in NGC~5128 can be
compared with period distributions of LPVs in other stellar systems (see 
Fig.~\ref{compareP}).
In old and metal-rich Galactic globular clusters, Miras span a rather
narrow range
of periods $140 \la P \la 310$ (Feast et al.~\cite{feast+02}), while
local Miras have periods ranging
from $\sim 100$ to $\sim 550$ days (Whitelock et al.~\cite{whitelock+00}).
According to Whitelock et al.~(\cite{whitelock+91}) there is no significant
population of Miras with periods longer than $\sim 700$ days in the 
Galactic bulge. However, their
result is in disagreement with Harmon \& Gilmore
(\cite{harmon&gilmore88}) and Blommaert et al.~(\cite{blommaert+98}).
These last authors found some OH/IR variables with periods in excess of
2000 days in the Galactic bulge.  Glass et al.\
(\cite{glass+95}) and Schultheis \& Glass (\cite{schultheis+glass01})
have analysed Sgr I field and Baade's window in
the Galactic bulge. The average period of 70 LPVs in Sgr I field
is $P_{av}=333$~day. In the Galactic Centre field the average period of
LPVs is 427 days, while that of the OH/IR stars amounts to 524 days
(Glass et al.~\cite{glass+01}). Apparently, longer-period, hence younger, LPVs
are concentrated more towards the Galactic Centre, where a mix of young and
old populations is observed due to a recent ($\la 1$~Gyr old) starburst.

Glass et al.~(\cite{glass+95}) discussed the
dependence of the type of LPVs discovered on the technique, in the sense that
optical searches found
principally shorter period Miras, while infrared techniques
(IRAS and OH/IR sources) preferentially found longer period Miras.
In our near-IR search, most of the short and longer period Miras
should be discovered,
with the exception of the most dust enshrouded OH/IR sources,
if these were present. We find in our fields as well that the periods of the
redder Miras are typically longer as shown by the shaded histogram in 
Fig.~\ref{compareP}. Unfortunately due to rather low statistics it is 
not possible at this moment to make similar comparisons in other systems.

In the LMC LPVs (including Mira and SR variables)
span a range of periods from about 20 to $\sim1000$ days
(e.g.\ Wood~\cite{wood00}, Cioni et al.~\cite{cioni+01}, Kiss \& Bedding
\cite{kiss&bedding03}).

It has been now quite firmly established that the period of Mira variables
represents a good indicator of the stellar population to which it
belongs (Feast \& Whitelock \cite{feast+whitelock87}).
Longer period Miras are expected to have
higher mass progenitors and therefore belong to a younger population
(e.g. Iben \& Renzini \cite{iben+renzini83}, Jura \& Kleinmann
\cite{jura&kleinmann92a,jura&kleinmann92b}, Kerschbaum \& Hron
\cite{kerschbaum&hron92}, Feast \& Whitelock \cite{feast+whitelock00}).

\begin{figure}
\centering
\includegraphics[width=7cm,angle=270]{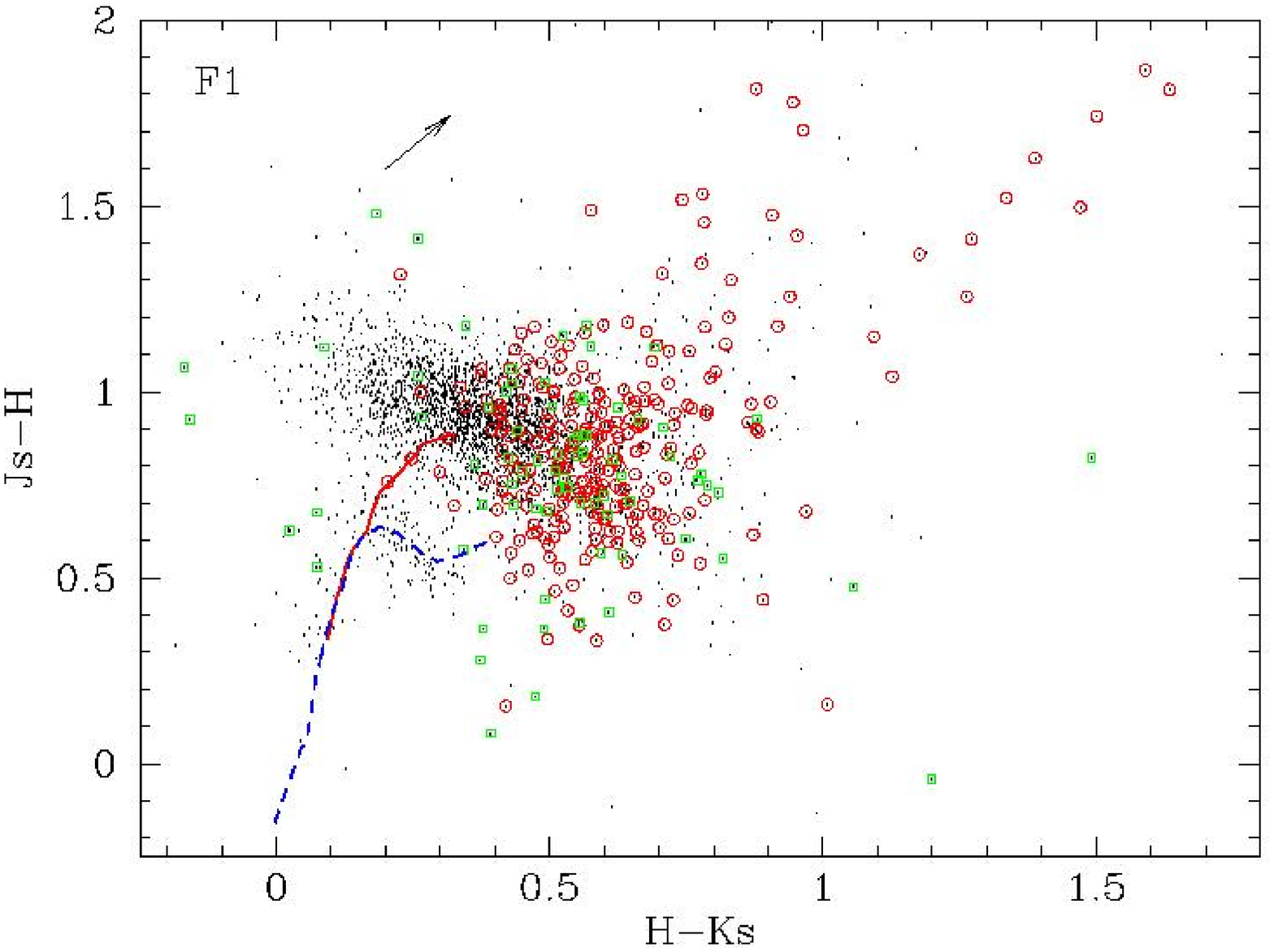}\\
\includegraphics[width=7cm,angle=270]{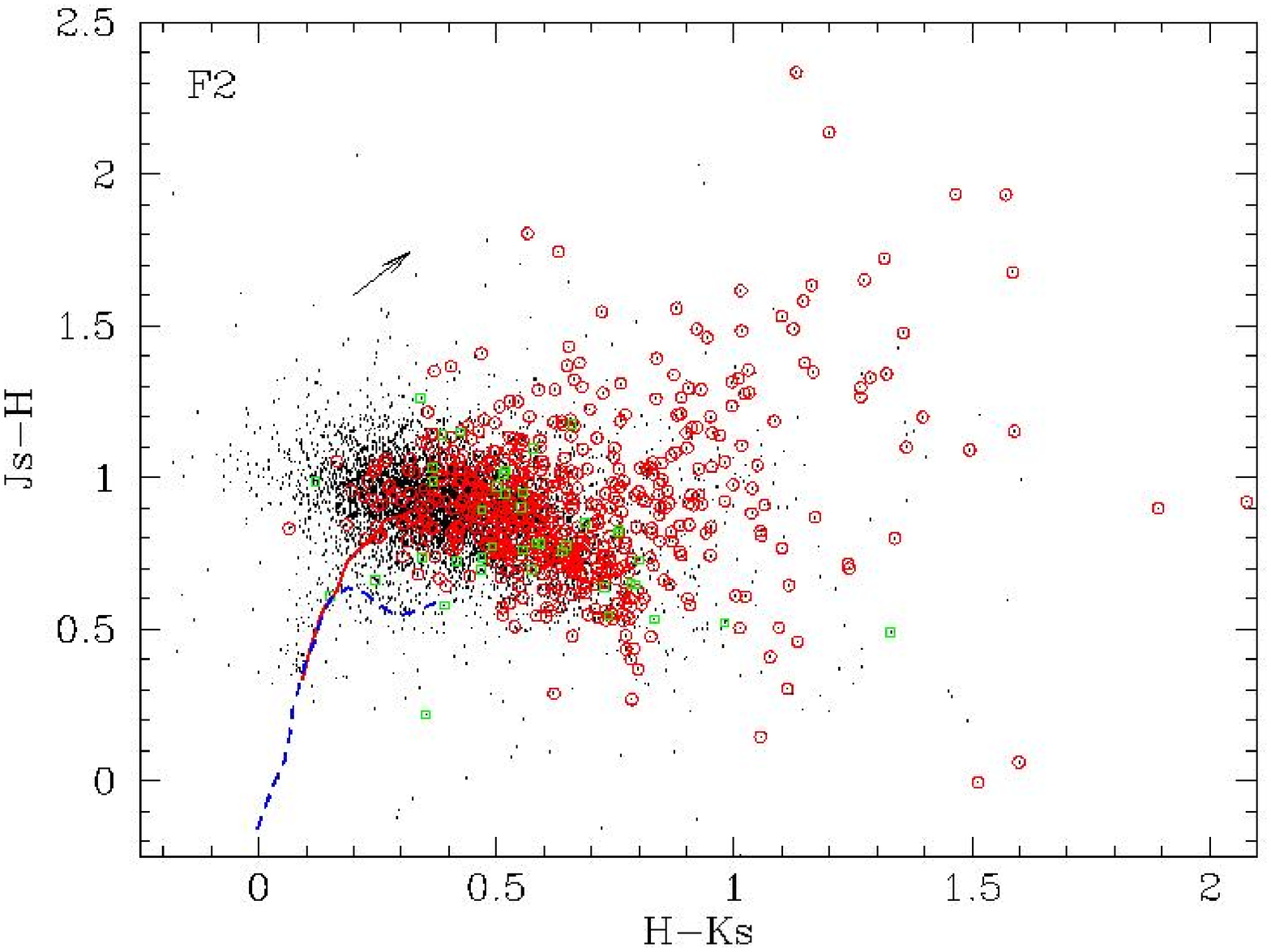}
\caption[]
	{Color-color diagrams for all the stars with ALLFRAME photometry errors
	smaller than 0.25. LPVs are indicated with
	red circles (secure periods) and green squares (less certain periods).
	Reddening vector for E(B-V)=0.5 is shown. Lines indicate intrinsic
	colors of Galactic giants (solid red line) and
	dwarfs (dashed blue line) from Bessell \&
	Brett (\cite{bessell+brett88}).
	}
\label{color-color}
\end{figure}

Unfortunately a precise empirical calibration of the initial mass
or, equivalently, the age vs.\ period of Miras is
lacking. It can be either inferred from the kinematical
properties of Miras in the solar neighbourhood (Feast \cite{feast63},
Feast \& Whitelock \cite{feast+whitelock00}),
or from the observations of Miras in simple stellar populations like
star clusters. According to Feast \& Whitelock (\cite{feast+whitelock87}),
the kinematic age of a 400~day Mira is of the order of $\sim 5$
Gyr and an 850 day period Mira would be younger than about 4~Gyr. Feast
(\cite{feast96}) on the other hand explored the metallicity dependence of the
mass-period relation for Miras and suggested that a 400~day Mira with 
a solar metallicity would have a mass of $\sim 1.4~M_\odot$. 
According to models (Iben \& Renzini \cite{iben+renzini83},
Vassiliadis \& Wood  \cite{vassiliadis+wood93}) a 400 day
Mira would have initial mass of about $\sim 1.0~M_\odot$.
However, it should be remembered that the age of the AGB star
with a $1.0~M_\odot$ progenitor will strongly depend on metallicity.
The more metal-rich, the older the star, assuming a constant mass. 
For example,
the turn-off age for a Z=0.004 star of $1.0~M_\odot$ is $\sim 4.6$~Gyr,
while a same mass solar-metallicity turn-off star would be 7.7 Gyr old
(Pietrinferni et al.~2003, in preparation). Turn-off age for a solar 
metallicity star of $1.4~M_\odot$ is less than 3~Gyr.

Miras in the old and metal-rich globular clusters have ages of the
order of 10--12 Gyr, with corresponding masses of the order of
$0.6-0.7 M_\odot$ and periods around $\sim 250$ days. Another empirical
point, calibrating the mass-period relation is given by Nishida
et al.~(\cite{nishida+00}), who have determined periods of 3 Miras in
the LMC star clusters with ages of 1.6 to 2.0 Gyr. All three Miras have
similar periods ranging from 491 to 528 days and they follow the
Mira PL relations determined from shorter-period stars
(Feast et al.~\cite{feast+89}).

NGC~5128 has a similar period distribution to Galactic bulge Miras, but
there is a tail with some $\sim 10\%$ of the Miras having periods
longer than 500 days. Unless this is purely due to
age-metallicity-period degeneracy, we conclude that this is an evidence for 
an intermediate-age AGB component in the NGC~5128 halo.

\section{Are there C-stars in NGC~5128?}

\subsection{Color-color diagram}
\label{color-color-sect}

\begin{figure}
\centering
\includegraphics[width=7cm,angle=270]{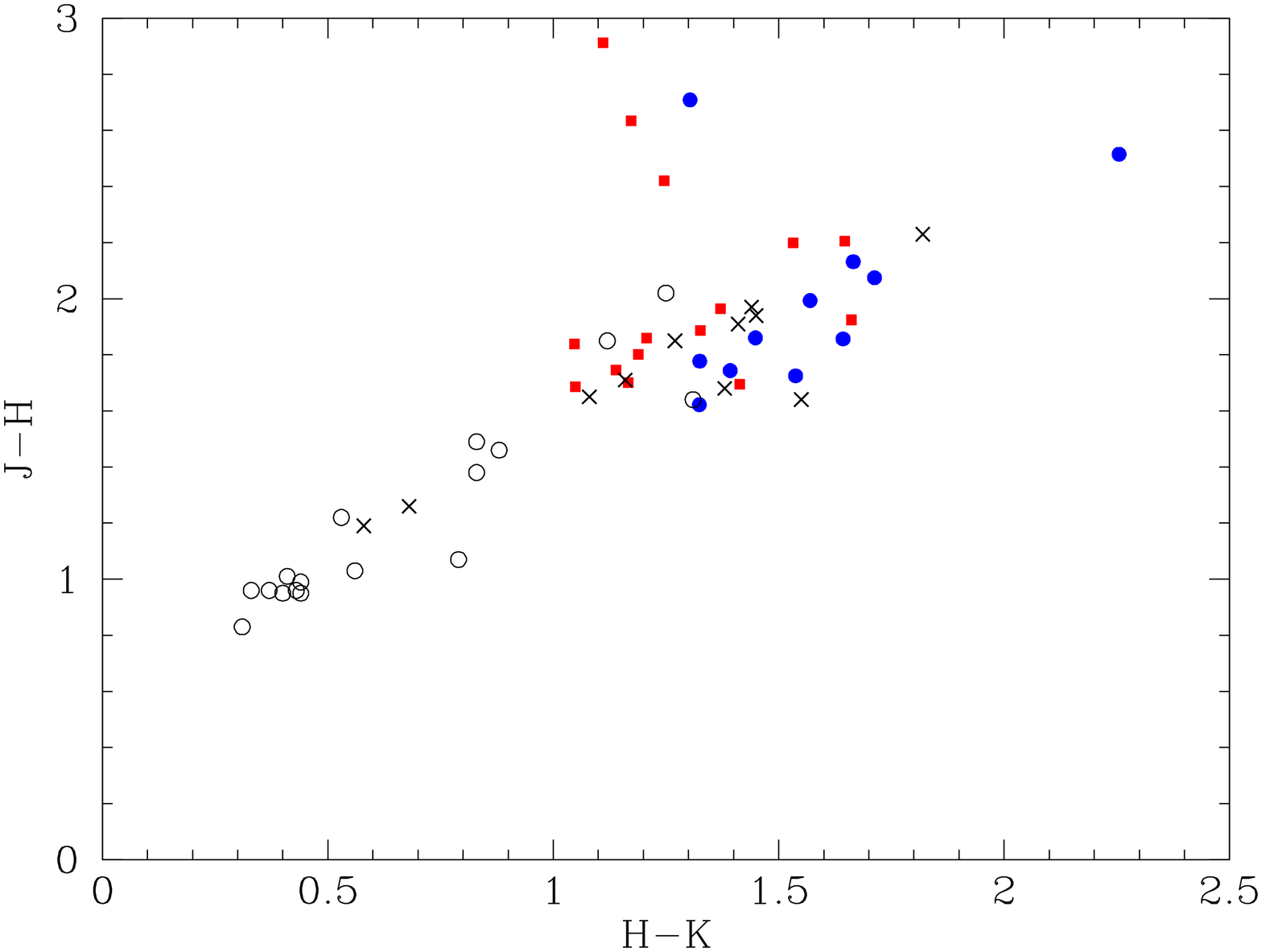}\\
\includegraphics[width=7cm,angle=270]{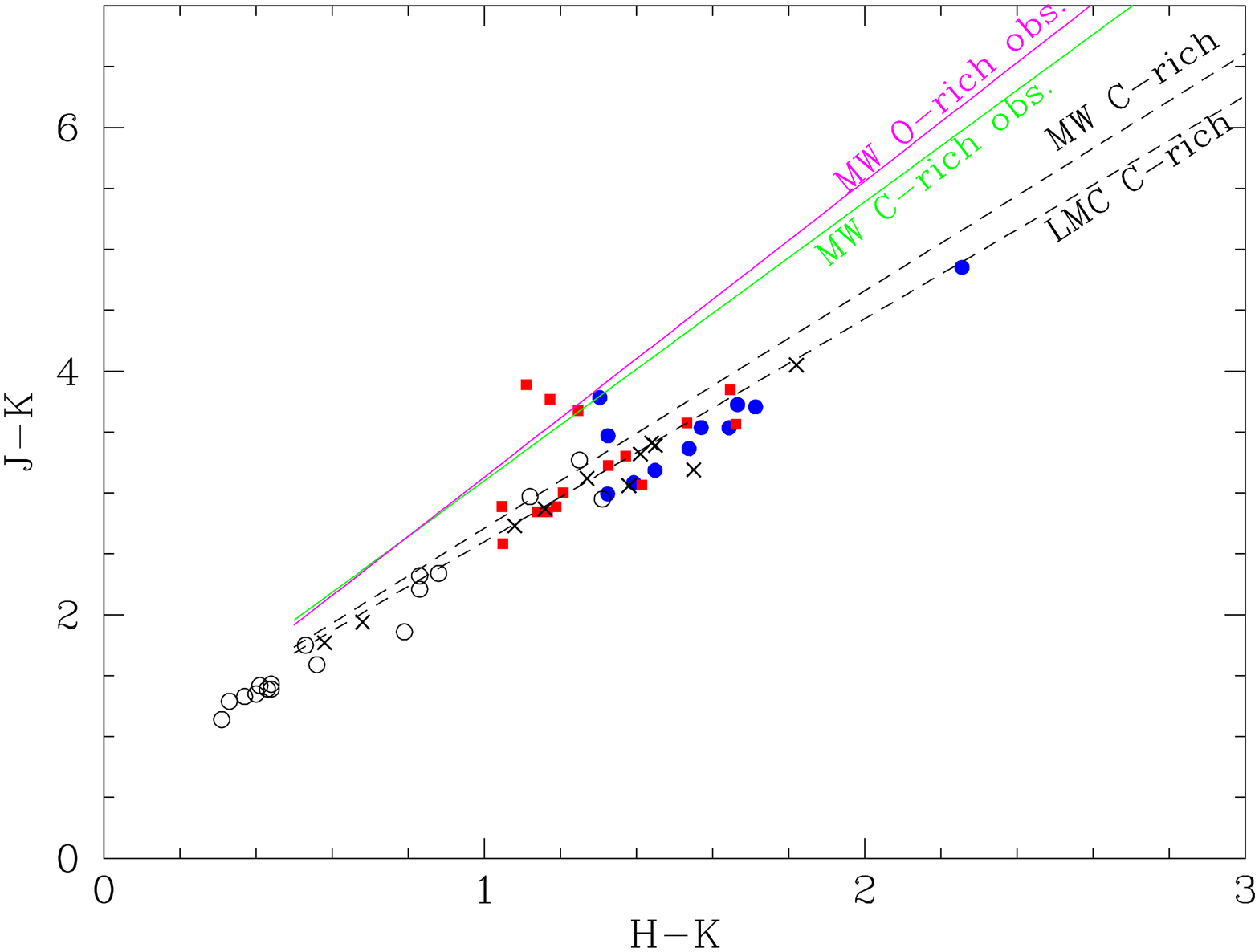}
\caption[]{Bona fide C-star candidates in NGC~5128:
	LPVs with $H-K \ga 1.0$ and $J-H \ga 1.5$ in NGC~5128
	are plotted as filled blue circles (Field~1) and filled red
	squares
	(Field~2). Obscured O-rich (open circles) and C-rich
	(crosses) LPVs in the LMC (Whitelock et al.~\cite{whitelock+03})
	are plotted for comparison.
	See text for the explanation of the model colors.}
\label{cstars}
\end{figure}

Color-color diagrams for all the stars with ALLFRAME photometry errors smaller
than 0.25 are shown in Fig.~\ref{color-color}.
Magnitudes have been de-reddened
adopting a foreground reddening of
$\mathrm{E}(\mathrm{B}-\mathrm{V})=0.11$ and
a Cardelli et al.~\cite{cardelli+89} reddening law. A reddening vector
corresponding to $\mathrm{E}(\mathrm{B}-\mathrm{V})=0.5$ is plotted.
As for Fig.~\ref{cmdlpv}, single-epoch $K$-band magnitudes closest
in time to $J$ and $H$-band measurements are used to form colors.
The solid (red) line is a Bessell \& Brett (\cite{bessell+brett88})
fiducial to the Galactic giants and the dashed (blue) line is for dwarfs.
Bessell \& Brett filters have been transformed to the photometric system of
ISAAC (Chris Lidman, private communication).

\begin{figure}
\centering
\includegraphics[width=7cm,angle=270]{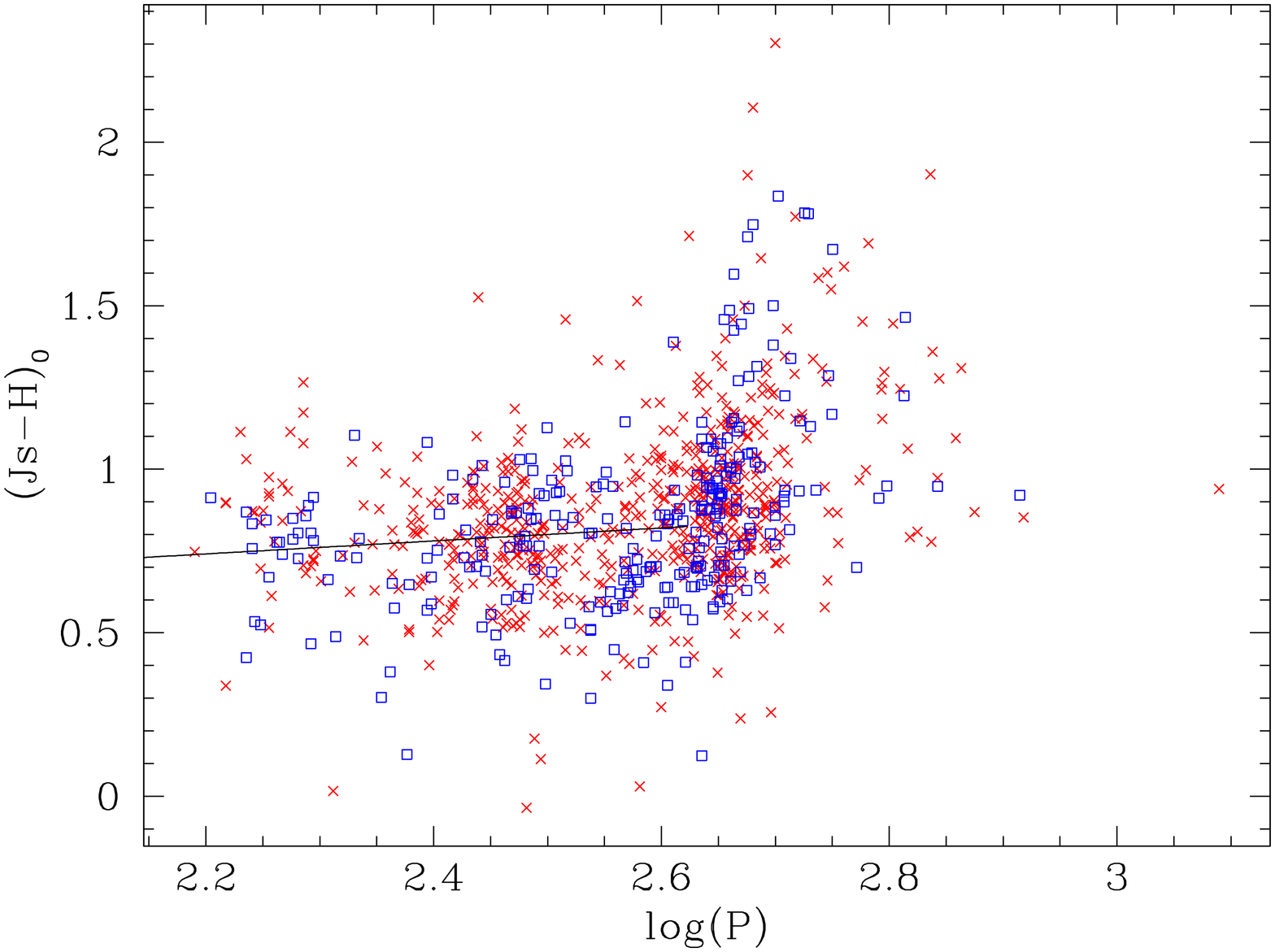}\\
\includegraphics[width=7cm,angle=270]{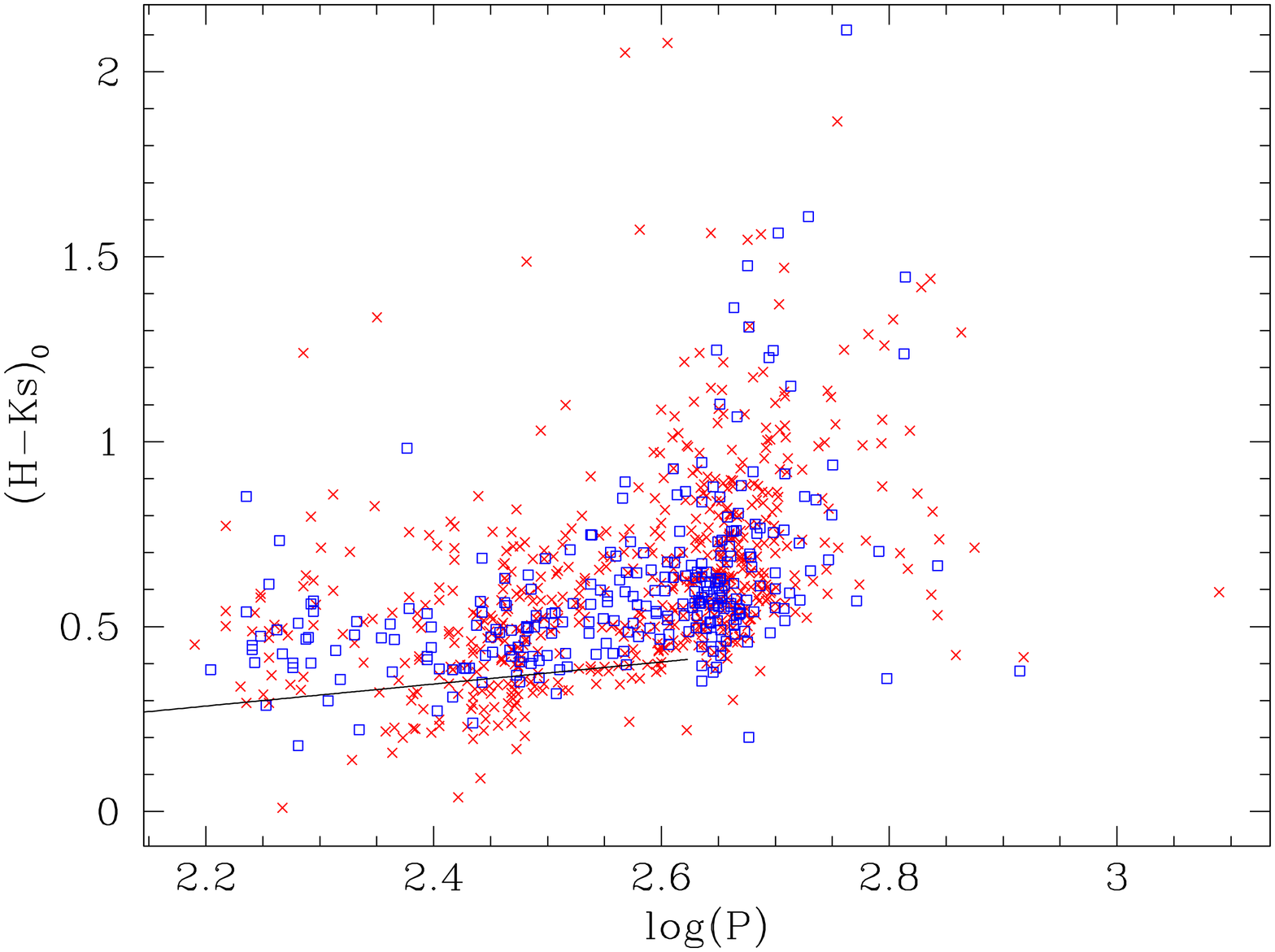}\\
\includegraphics[width=7cm,angle=270]{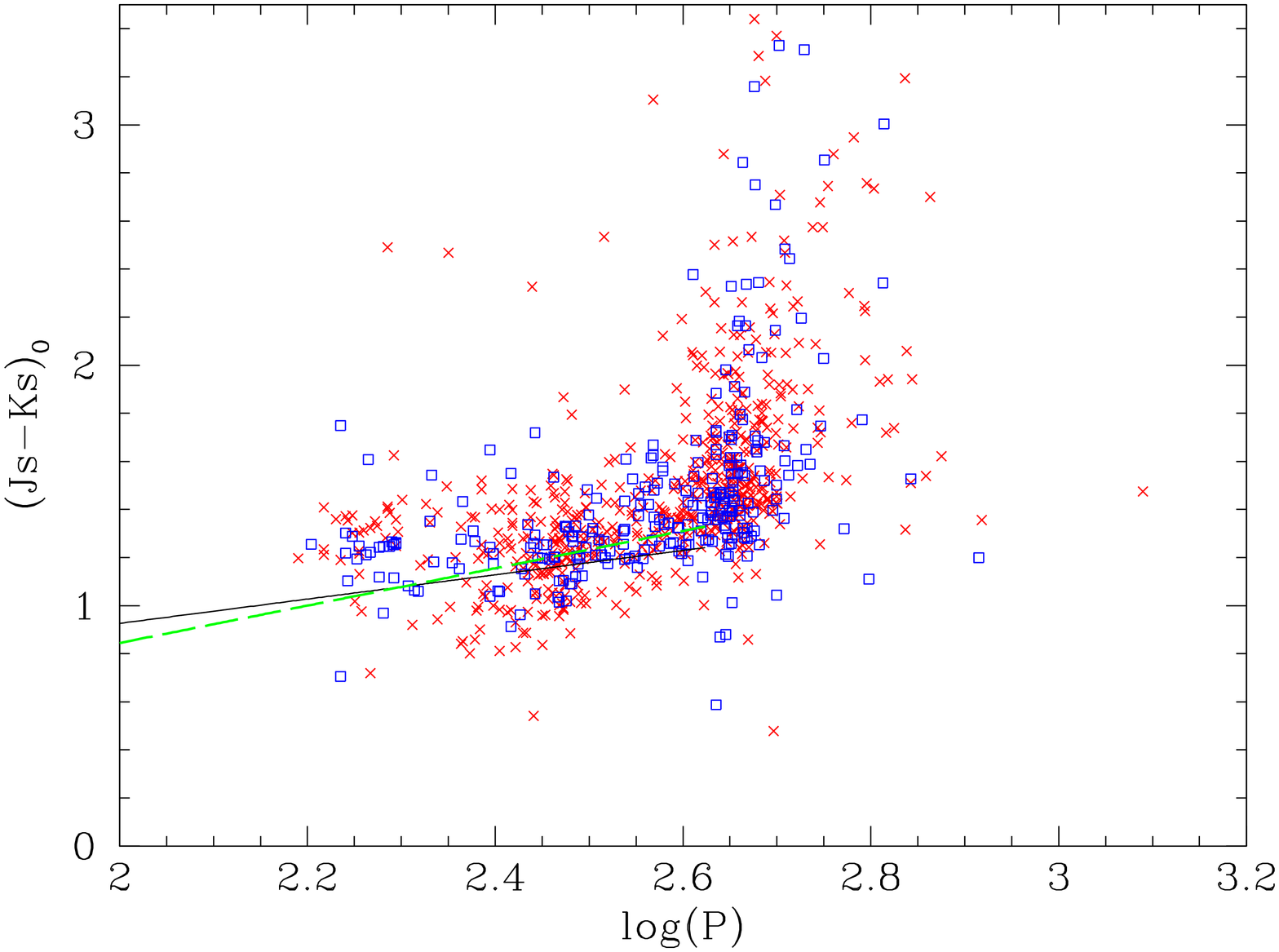}
\caption[]
	{Period-color relations. For comparison a linear fit to the LMC
	period-color relations (Feast et al.~\cite{feast+89}) is shown 
	(solid line in all three panels). In period-$(J-K)$ diagram
	a linear fit to Sgr~I Miras (Glass et al.~\cite{glass+95})
	is shown with dashed line.	
	}
\label{Pcol}
\end{figure}

Carbon stars are typically found among intermediate-age metal-poor
populations. Single stars enter the AGB phase as oxygen-rich (O-rich)
and they can be converted to carbon-rich (C-rich) after the products
of nuclear burning are brought up to the surface during thermal
pulses. Due to the lower oxygen abundance of a metal-poor star, fewer
C-atoms are needed to convert it into a C-rich star than is the case
for a more metal-rich star.  The ratio of the number of carbon-rich
stars to oxygen-rich stars (C/M ratio) can therefore be used to derive
metallicity gradients (Cioni \& Habing, \cite{cioni+habing03}).
Detecting carbon stars in NGC~5128 is important because their presence
would be a definite proof of an intermediate-age and metal-poor
component.  Their origin might belong to a recently accreted
metal-poor, LMC-like companion galaxy.

Carbon stars in the LMC have $1.5<J-K_s<2.0$. Stars redder than $J-K_s>2.0$
are obscured and can either be O-rich or C-rich (Cioni et
al.~\cite{cioni+01}).
In the $H-K$ vs. $J-H$ color-color diagram, carbon stars are located
redwards of
$H-K \ga 0.4$ and $J-H \ga0.8$. However, O-rich LPVs can also be found there.
According to Bessell \& Brett  (\cite{bessell+brett88}) there should be no
O-rich LPVs redder than $H-K \ga 1.0$ and $J-H \ga 1.4$
(see their Fig.~A3). There are 11 stars in Field~1 and 15 in Field~2
with $H-K$ and $J-H$ colors consistent with being carbon stars
and all but three of them have periods longer
than 450 days and the average period of these stars is 526 days. Their
mean amplitude is 1.2~mag. All of them have $J_s-K_s>2.5$. Very red
$J_s-K_s$ colors and large amplitudes indicate that they probably have
circumstellar shells.

A comparison of these stars with obscured C-rich and O-rich AGB variables in the
LMC (Whitelock et al.~\cite{whitelock+03}) is shown in Fig.~\ref{cstars}.
The LMC O-rich AGB stars are plotted with open circles
and the C-rich stars with crosses. NGC~5128 candidate C-rich LPVs are
plotted as filled blue circles (Field~1) and filled red squares (Field~2).
In order to facilitate the comparison, ISAAC data have been
transformed into the SAAO photometric system.
Cohen et al.~(\cite{cohen+81}) have shown that the
$(H-K)$ and $(J-H)$ colors of optically visible C-rich stars in the Milky
Way are correlated due to line blanketing by molecular absorption bands
(shown as dashed line labelled ``MW C-rich'').
The correlation for the optically visible C-rich stars in the LMC
(dashed line labelled ``LMC C-rich'') was published by Costa \&
Frogel (\cite{costa+frogel96}), while the correlations for the obscured
C-rich (green solid line labelled ``MW C-rich'') and O-rich (magenta solid
line labelled ``MW O-rich'') Milky Way stars comes from Guglielmo et al.\
(\cite{guglielmo+93}).
All four correlation equations are taken from van Loon (\cite{vanloonPhD}),
who transformed them to $(H-K)$ and $(J-K)$ on the SAAO system.
Stars lying along the correlations for obscured O-rich and C-rich stars could
belong to either category, but the majority lies along the lines where
optically visible C-stars are found. These are bona fide C-star candidates.
However, more C-stars could be present among the long-period stars with
somewhat bluer colors, but without spectroscopy they cannot be
distinguished from the obscured O-rich LPVs.

\subsection{Period-color distributions}

Color-log period diagrams for NGC~5128 LPVs are shown in Fig.~\ref{Pcol}.
For comparison, a linear fit to the LMC
color-log period relations (Feast et al.~\cite{feast+89}) are shown as
solid lines. They have been shifted to the ISAAC photometric system.
It should be noted that these linear relations are valid only for
$P<420$~day. As for
Figs.~\ref{cmdlpv} and \ref{color-color}, single-epoch $K$-band
magnitudes closest in time to
$J$ and $H$-band measurements were used to form colors.
This produces a large scatter around the mean Period-color relation.
The scatter is the largest in $J-H$ vs.\
log(P) diagram where the phase difference between the observations is
larger and thus some colors might even be non-physical.

$(J-H)_0$ and $(J-K)_0$ colors are likely to be affected differently by
metallicity changes, but also by gravity and atmospheric extension.
Whitelock et al.~(\cite{whitelock+91}) have compared
$(J-H)_0$ and $(J-K)_0$ vs. $\log P$ diagrams for the LMC, the Galactic
globular clusters, the solar neighbourhood and the Galactic bulge.
Their conclusion is that at a given period, colors of Miras in these different
environments are very similar. However, Miras in the Sgr I field of
the Galactic bulge (Glass et al.~\cite{glass+95}) have redder
$(H-K)_0$ and slightly bluer $(J-H)_0$ colors than the LMC Miras
at a given period, while $(J-K)_0$ colors do not show so large offset. 
Feast~(\cite{feast96}) reports the mean $(J-K)_0$ 
color-period relation 
for Sgr~I Miras studied by Glass et al.\ (\cite{glass+95}) which has
quite a different slope with respect to the LMC relation. He ascribes the
difference to metallicity differences. The SgrI relation is shown in
the $(J-K)_0$ color--log~P diagram as a long dashed line. It provides 
much better fit to NGC~5128 LPVs and suggests a similar metallicity as in 
the Galactic bulge. The large scatter around the color-period relation
due to random phase $J_s$-band observations and possible abundance 
spread within NGC~5128 halo (e.g.\ Harris et al.~\cite{harris+99}, 
Rejkuba et al.~\cite{rejkuba+01}) prevents us from drawing more
quantitative comparison. 

Some very red stars are present among the longest period variables in
NGC~5128.  This has also been seen in the LMC, the solar neighbourhood
and the Bulge (Glass et al.~\cite{glass+95}, Kiss \& Bedding 
\cite{kiss&bedding03}), and is ascribed to the
presence of cool circumstellar shells arising from mass loss. Such
shells contribute principally at longer wavelengths creating an
infrared excess.  Actually there are 113 sources in Field~1 and 236 in
Field~2 with $K_s$-band magnitudes brighter than the RGB tip and no $J_s$ or
$H$-band counterparts. Many of these could be dust enshrouded LPVs
similar to those found in the central parts of the Milky Way by van
Loon et al.\ (\cite{vanloon+03}).  Unfortunately, periods could only be
measured for a handful of them.

%
%

\section{Conclusions}

We have analysed near-IR properties of 897 LPVs with well 
determined periods and $J_sHK_s$ photometry in two halo fields in NGC~5128. 
Mostly they are found to be brighter than the tip of the 
RGB ($K_s<21.24$; Rejkuba \cite{rejkuba03}) with magnitudes
ranging from about $K=19$ to $K=21.5$. They have periods
between 155 and 1000 days and K-band amplitudes between 0.1 and 2 mag,
characteristic of semi-regular and Mira variables.
They obey the same period-luminosity relation as found in the Magellanic
Clouds, Solar neighbourhood and the Galactic bulge 
(Rejkuba \cite{rejkuba03}). Here we have
compared the colors, periods and amplitudes of these variables with those
found in old stellar populations like Galactic globular clusters and
Galactic bulge as well as with intermediate-age Magellanic Cloud long period
variables.

There is an excess of short period, globular cluster-like, 
Mira variables in Field~2 showing the differences between 
the stellar populations of the two fields. The brightness of the 
AGB tip, which occurs around $M_K=-8.65$ in NGC~5128, 
corresponding to $M_{bol}= -5.3$, compared to $M_K=-8$ in the 
Galactic bulge (Zoccali et al.~\cite{zoccali+03}) and in the M31 bulge, 
(Stephens et al.~\cite{stephens+03}) is an evidence for a 
$\sim 4$~Gyr old component (Mould \& Aaronson \cite{mould&aaronson82}). 
Some 10\% 
of the long period $P>500$~day Miras are expected to have ages younger 
than 7 Gyr and could be as young as $3-4$ Gyr old.
Approximately two dozen C-star candidates are identified, 
but more could be present among
the reddened long period variables. C-stars are expected to be found only 
among metal-poor AGB stars. 

%
%

\begin{acknowledgements}

We thank Chris Lidman for providing the transformation equations for
ISAAC filters and Santi Cassissi for providing the stellar evolutionary
tracks used to estimate ages. MR thanks M.-R. Cioni and A. Renzini for some
interesting discussions.
DM is sponsored by FONDAP Center for Astrophysics 15010003. 
TRB is grateful to the
Australian Research Council and P. Universidad Cat\'olica for financial
support and to the DITAC International Science \& Technology Program.
\end{acknowledgements}

\end{document}